\documentclass[conference]{IEEEtran}
\usepackage{caption}
\usepackage{algpseudocode}
\usepackage{amsmath}
\usepackage{amsfonts}
\usepackage{amssymb}
\usepackage{subfigure}
\usepackage{graphicx}
\usepackage[table]{xcolor}
\usepackage{array,multirow,graphics}

\usepackage{amsfonts}
\usepackage{booktabs}
\usepackage{siunitx}

\usepackage{color}

\usepackage{tikz}

\usepackage{enumitem}

\usepackage{url}
\makeatletter
\def\url@leostyle{%
  \@ifundefined{selectfont}{\def\UrlFont{\sf}}{\def\UrlFont{\small\ttfamily}}}
\makeatother
\newcommand{\eat}[1]{}
\usepackage[linesnumbered,ruled]{algorithm2e}
\usepackage{mdframed} 

\usepackage{xcolor}
\definecolor{light-gray}{gray}{0.9}

\newenvironment{packed_enum}{%
  \begin{enumerate}%
  }{\end{enumerate}}

\usepackage{amsthm}

\newcolumntype{L}[1]{>{\raggedright\let\newline\\\arraybackslash\hspace{0pt}}m{#1}}

\usepackage{listings}

\definecolor{codegreen}{rgb}{0,0.6,0}
\definecolor{codegray}{rgb}{0.5,0.5,0.5}
\definecolor{codepurple}{rgb}{0.58,0,0.82}
\definecolor{backcolour}{rgb}{0.95,0.95,0.92}
 
\lstdefinestyle{mystyle}{
    backgroundcolor=\color{backcolour},   
    commentstyle=\color{codegreen},
    keywordstyle=\color{magenta},
    numberstyle=\tiny\color{codegray},
    stringstyle=\color{codepurple},
    basicstyle=\footnotesize,
    breakatwhitespace=false,         
    breaklines=true,                 
    captionpos=b,                    
    keepspaces=true,                 
    numbers=left,                    
    numbersep=5pt,                  
    showspaces=false,                
    showstringspaces=false,
    showtabs=false,                  
    tabsize=2
}
 
\usepackage{listings, xcolor}

\definecolor{verylightgray}{rgb}{.97,.97,.97}

\lstdefinelanguage{Solidity}{
	keywords=[1]{anonymous, assembly, assert, balance, break, call, callcode, case, catch, class, constant, continue, constructor, contract, debugger, default, delegatecall, delete, do, else, emit, event, experimental, export, external, false, finally, for, function, gas, if, implements, import, in, indexed, instanceof, interface, internal, is, length, library, log0, log1, log2, log3, log4, memory, modifier, new, payable, pragma, private, protected, public, pure, push, require, return, returns, revert, selfdestruct, send, solidity, storage, struct, suicide, super, switch, then, this, throw, transfer, true, try, typeof, using, value, view, while, with, addmod, ecrecover, keccak256, mulmod, ripemd160, sha256, sha3}, 
	keywordstyle=[1]\color{blue}\bfseries,
	keywords=[2]{address, bool, byte, bytes, bytes1, bytes2, bytes3, bytes4, bytes5, bytes6, bytes7, bytes8, bytes9, bytes10, bytes11, bytes12, bytes13, bytes14, bytes15, bytes16, bytes17, bytes18, bytes19, bytes20, bytes21, bytes22, bytes23, bytes24, bytes25, bytes26, bytes27, bytes28, bytes29, bytes30, bytes31, bytes32, enum, int, int8, int16, int24, int32, int40, int48, int56, int64, int72, int80, int88, int96, int104, int112, int120, int128, int136, int144, int152, int160, int168, int176, int184, int192, int200, int208, int216, int224, int232, int240, int248, int256, mapping, string, uint, uint8, uint16, uint24, uint32, uint40, uint48, uint56, uint64, uint72, uint80, uint88, uint96, uint104, uint112, uint120, uint128, uint136, uint144, uint152, uint160, uint168, uint176, uint184, uint192, uint200, uint208, uint216, uint224, uint232, uint240, uint248, uint256, var, void, ether, finney, szabo, wei, days, hours, minutes, seconds, weeks, years},	
	keywordstyle=[2]\color{teal}\bfseries,
	keywords=[3]{block, blockhash, coinbase, difficulty, gaslimit, number, timestamp, msg, data, gas, sender, sig, value, now, tx, gasprice, origin},	
	keywordstyle=[3]\color{violet}\bfseries,
	identifierstyle=\color{black},
	sensitive=false,
	comment=[l]{//},
	morecomment=[s]{/*}{*/},
	commentstyle=\color{gray}\ttfamily,
	stringstyle=\color{red}\ttfamily,
	morestring=[b]',
	morestring=[b]"
}

\begin{document}

\title{Scalable and Privacy-preserving Design of On/Off-chain Smart Contracts}

\author{
{Chao Li, Balaji Palanisamy and Runhua Xu}%
\vspace{1.6mm}\\
\fontsize{10}{10}\selectfont\itshape
School of Computing and Information, University of Pittsburgh, USA
\vspace{1.6mm}\\
\{chl205, bpalan, runhua.xu\}@pitt.edu\\
\vspace{1.2mm}\\
\fontsize{10}{10}\selectfont\rmfamily\itshape
\vspace{-0.55in}
}%

\maketitle

\begin{abstract}
The rise of smart contract systems such as Ethereum has resulted in a proliferation of blockchain-based decentralized applications including applications that store and manage a wide range of data. 
Current smart contracts are designed to be executed solely by miners and are revealed entirely \textit{on-chain}, resulting in reduced scalability and privacy.
In this paper, we discuss that scalability and privacy of smart contracts can be enhanced by splitting a given contract into an \textit{off-chain} contract and an \textit{on-chain} contract.
Specifically, functions of the contract that involve high-cost computation or sensitive information can be split and included as the \textit{off-chain} contract, that is signed and executed by only the interested participants.
The proposed approach allows the participants to reach unanimous agreement \textit{off-chain} when all of them are honest, allowing computing resources of miners to be saved and content of the \textit{off-chain} contract to be hidden from the public.
In case of a dispute caused by any dishonest participants, a signed copy of the \textit{off-chain} contract can be revealed so that a verified instance can be created to make miners enforce the true execution result.
Thus, honest participants have the ability to redress and penalize any fraudulent or dishonest behavior, which incentivizes all participants to honestly follow the agreed \textit{off-chain} contract.
We discuss techniques for splitting a contract into a pair of \textit{on/off-chain} contracts and propose a mechanism to address the challenges of handling dishonest participants in the system.
Our implementation and evaluation of the proposed approach using an example smart contract demonstrate the effectiveness of the proposed approach in Ethereum.
\end{abstract}

\section{Introduction}
\label{s1}
Creating trust among mutually distrustful participants without involving a trusted third party has been a challenge for several decades. 
The growth of Bitcoin~\cite{nakamoto2008bitcoin} and the emerging follow-up cryptocurrencies have positioned blockchain as a promising solution for creating trust in a decentralized environment.
Smart contracts expand the use of blockchains by allowing mutually distrustful participants to reach an agreement upon the execution results of complex contracts without a trusted third party.
In leading smart contract systems such as Ethereum~\cite{wood2014ethereum} and NEO~\cite{NEO}, every single step of the decentralized computation is performed and verified by miners in the blockchain network, who are incentivized by the cryptocurrency rewards to behave honestly.
This combination of decentralized computation and cryptocurrency-based incentives has led to the development of a large number of decentralized applications including applications that store and manage a wide range of data~\cite{DAPP}.
Ethereum~\cite{wood2014ethereum}, the first blockchain system that supports Turing-complete smart contracts, had a peak market cap of \$134 billion~\cite{ETH} in 2018.
The Smart Contracts market is estimated to grow at a CAGR of 32\% during the period 2017 to 2023~\cite{marketresearchfuture}.

Current smart contracts are designed to be executed solely by miners and are revealed entirely \textit{on-chain}, resulting in reduced scalability and privacy.
In this paper, we discuss that scalability and privacy of smart contracts can be enhanced by splitting a contract into two separate contracts:
(1) an \textit{off-chain} contract encapsulating functions of the whole contract that involve high-cost computation and/or distinguishable logic that may reveal private information about the participants; 
(2) an \textit{on-chain} contract encapsulating the remaining low-cost/non-sensitive functions of the whole contract.
Replacing the whole smart contract with the pair of on/off-chain contracts enables deploying only the \textit{on-chain} contract onto the blockchain. 
This also allows saving the computing cost of running the \textit{off-chain} contract by miners in the Ethereum blockchain network while the sensitive information involved in the \textit{off-chain} contract can be hidden from the public.

After generating the pair of on/off-chain contracts from the whole contract, 
when all the participants are honest, they can reach unanimous agreement on the result of \textit{off-chain} contract, just as if the \textit{off-chain} contract was deployed \textit{on-chain} and executed by miners.
However, if there is a dispute of the \textit{off-chain} result, such as a dishonest participant trying to lie about it, it would be necessary to allow honest participants to resolve the dispute \textit{on-chain}.
In this paper, we propose an effective mechanism that allows any honest participant to leverage the already deployed \textit{on-chain} contract to create an \textit{on-chain} instance of the \textit{off-chain} contract.
The process of creating a new contract from an existing contract can establish a unique connection between the two contracts.
Through this connection, only the \textit{off-chain} contract agreed and signed by all the participants can pass the integrity verification and can be created by the \textit{on-chain} contract as a verified instance.
Then, only the execution result of the verified instance can enforce the state change of the \textit{on-chain} contract.
Thus, honest participants have the ability to redress and penalize any fraudulent or dishonest behavior, which incentivizes all participants to honestly follow the agreed \textit{off-chain} contract.
We implement and evaluate the proposed approach in the Ethereum official test network \textit{Kovan}~\cite{Kovan2019} with an example smart contract using \textit{Solidity}~\cite{Solidity2017}.
Our implementation demonstrates the effectiveness of the proposed approach of using \textit{on/off-chain} contracts in Ethereum.

\section{On/off-chain smart contracts}
\label{s2}

In this section, we first introduce the creation and execution of a general smart contract in Ethereum.
We then present the techniques for splitting a given smart contract into a pair of \textit{on/off-chain} contracts and explain how the pair of contracts can be executed in Ethereum.


\subsection{Smart contracts in Ethereum}
\label{s2-1}

There are two types of accounts in Ethereum, namely External Owned Accounts (EOAs) controlled by private keys owned by users and Contract Accounts (CAs) assigned to smart contracts.
A user of Ethereum should first create a EOA with a pair of keys and then deploy smart contracts from the EOA, resulting in the creation of CAs associated with the smart contracts.
A smart contract in Ethereum refers to a piece of program code that usually consists of multiple functions, a few parameters and perhaps some modifiers.
After programming a smart contract in a language such as \textit{Solidity}~\cite{Solidity2017}, a user can compile the contract to get its \textit{bytecode} and Application Binary Interface (\textit{ABI}) and can send a contract creation transaction to the Ethereum network with \textit{bytecode} and (optional) \textit{ABI}. 
Upon receiving the transaction, miners will include the \textit{bytecode} into the next block, meaning that a new smart contract has been created, whose CA can be deterministically computed from the address of its creator and a nonce.
Each CA can be viewed as a small decentralized server that can act based on the functions in its contract and can store data (e.g., cryptocurrency) allowed by its contract.
However, CAs are passive, meaning that execution of any function of deployed smart contracts must be invoked through either transactions sent by EOAs or messages sent from CAs.
As a result, the transactions/messages, as well as function inputs inside them, are all recorded by the Ethereum blockchain, which makes the function outputs deterministic because all miners can execute the function with the same inputs and gets the same outputs.
It is worth noting that one needs to pay Gas~\cite{wood2014ethereum} for either deploying a new smart contract or calling a function of existing smart contracts in Ethereum.
Gas can be exchanged with Ether, the cryptocurrency used in Ethereum, and Ether can be exchanged with real money. 

As discussed above, Ethereum requests each invoked function of deployed \textit{on-chain} contracts to be executed by all the miners for the purpose of getting trustworthy outputs in the decentralized environment.
However, this \textit{all-on-chain} execution model of smart contracts may drain the scarce network resources and introduce privacy risks, when the invoked function includes high-cost calculation or sensitive information.
Therefore, instead of deploying a whole contract with all function included, 
we propose to split such problem functions from the whole contract to execute them \textit{off-chain} while still executing the rest of functions \textit{on-chain}.
We present the technique of implementing this approach by splitting a given contract into a pair of \textit{on/off-chain} contracts.

\begin{figure}
\centering
{
    \includegraphics[width=8cm,height=7cm]{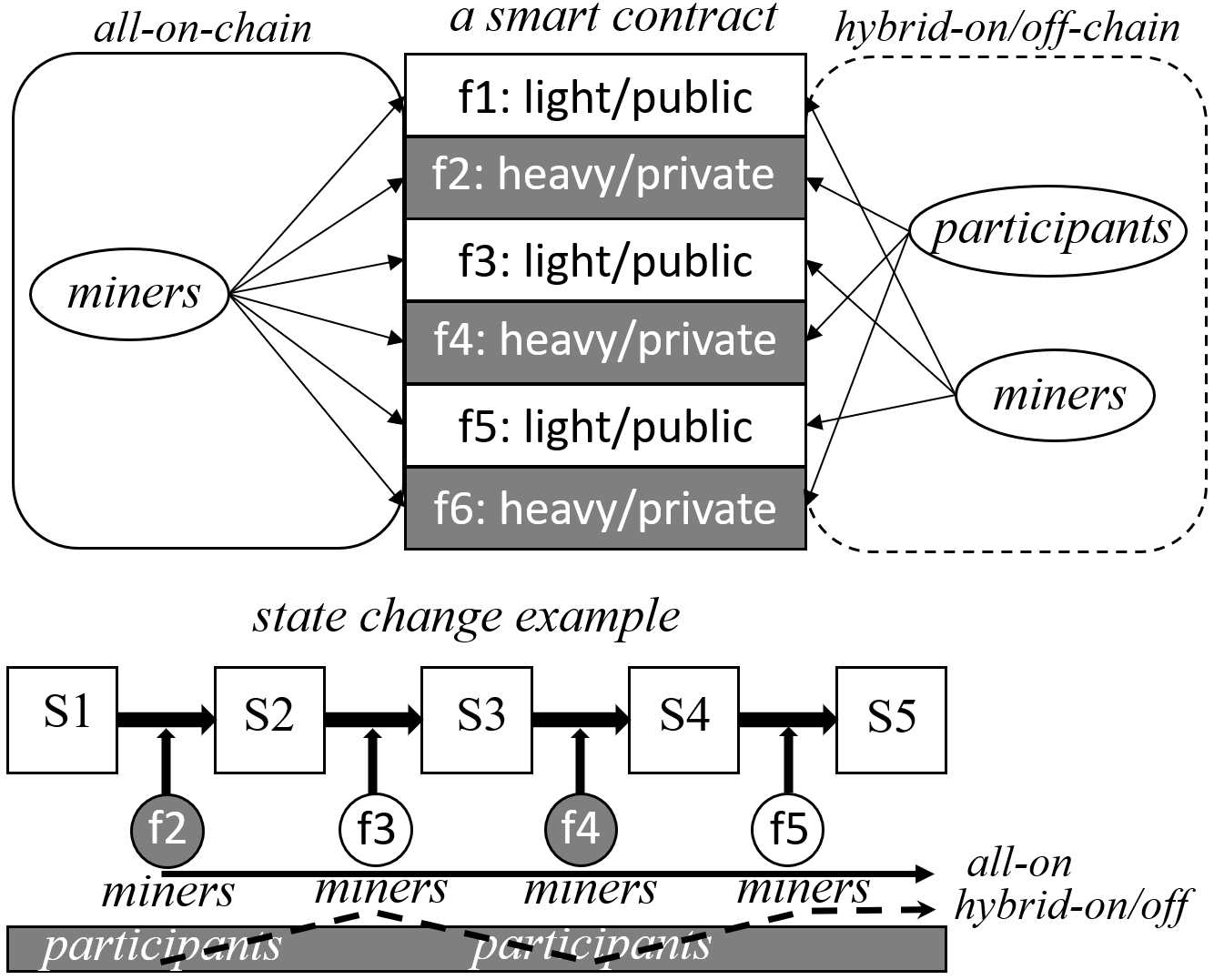}
}
\caption {\small Comparison of the current \textit{all-on-chain} execution model of smart contracts in Ethereum and the proposed \textit{hybrid-on/off-chain} execution model}
\label{f1} 
\end{figure}

\subsection{On/off-chain smart contracts}
\label{s2-2}

We broadly classify functions of smart contracts into two categories based on the computing resources spent for executing them and the sensitive information carried by them: (1) \textit{heavy/private} functions involving high-cost calculation or sensitive information and (2) \textit{light/public} functions involving none of these features.
For example, Alice and Bob may decide to bet on a private topic between them using cryptocurrency they have, so they draft a simplified betting contract as shown in Algorithm~\ref{a1}, which consists of three functions. They can first make deposits to the contract through \textit{deposit()}, then invoke \textit{reveal()} after a certain temporal threshold to reveal the result and finally reassign the cryptocurrency locked in the contract based on the result by calling \textit{reassign()}.
Since both \textit{deposit()} and \textit{reassign()} are simple cryptocurrency transfer functions, they contain neither high-cost calculation nor sensitive information, so they can be allocated to \textit{light/public} functions. In contrast, \textit{reveal()} may contain details of the customized betting rules that are private to the participants and may involve an arbitrary amount of computational cost, hence it should be allocated to a \textit{heavy/private} function.
It could be quite customized to allocate functions into the two categories. Here we recommend to allocate all functions of cryptocurrency transfer into \textit{light/public} functions and consider the remaining ones as \textit{heavy/private} functions.

\vspace{2mm}

\begin{minipage}{\linewidth}
\begin{lstlisting}[
linewidth=8.3cm,
language=Solidity,
basicstyle=\scriptsize,
caption= A simplified betting contract,
label={a1},
frame=single,
numbers=none,
numbersep=5pt,
breaklines=true,
breakatwhitespace=true,
]
contract betting {
  function deposit() payable public participantOnly;
  function reveal() public participantOnly;
  function reassign() public participantOnly;
}
\end{lstlisting}
\end{minipage}

After all functions of a smart contract has been classified, the \textit{heavy/private} functions and \textit{light/public} functions can be grouped into an \textit{off-chain} contract and an \textit{on-chain} contract, respectively.
Then, only the \textit{on-chain} contract needs to be deployed to be publicly executed by all the miners while the \textit{off-chain} contract can be privately executed by only a small group of interested participants.
Ideally, splitting the \textit{off-chain} contract from the whole contract does not affect the state change of the \textit{on-chain} contract.
For clarity of illustration, consider an example smart contract in Fig.~\ref{f1}, which consists of three light/public functions (i.e., $c1$, $c3$, $c5$) and three heavy/private functions (i.e., $c2$, $c4$, $c6$).
With the \textit{all-on-chain} model, the whole contract consisting of all the six functions is deployed and miners need to execute function $f2$, $f3$, $f4$ and $f5$ to change the state of the deployed contract from $S1$ to $S5$.
In contrast, with the \textit{hybrid-on/off-chain} model, only an \textit{on-chain} contract consisting of function $f1$, $f3$ and $f5$ is deployed, so miners only need to execute function $f3$ and $f5$ while $f2$ and $f4$ can be privately executed by interested participants.
As long as the participants are honest, they will be able to reach unanimous agreement on the \textit{off-chain} execution results of $f2$ and $f4$ and input these results to turn the contract state from $S1$ to $S2$ and later from $S3$ to $S4$, just as if both $f2$ and $f4$ were executed by miners \textit{on-chain}. However, in case of dishonest participants trying to lie about the \textit{off-chain} execution results, we need an additional mechanism to always allow honest participants to enforce the \textit{true} execution results of the \textit{off-chain} contract.
We will introduce this mechanism in the next section.

\section{Enforcing off-chain contracts}
\label{s3}

Our proposed enforcement mechanism is used by honest participants to enforce the results of \textit{off-chain} execution and penalize any dishonest participants. 

In the presence of any dishonest participants, an honest participant can simply deploy the \textit{heavy/private} functions back to the blockchain, make them get recomputed by the miners to enforce the state of \textit{on-chain} contract to be changed as expected.
However, there are two challenges to make this strategy work as expected.
First, the honest participant must prove that the deployed \textit{heavy/private} functions are exactly same as the original ones agreed by all the participants at the beginning. Otherwise, participants can falsify these functions based on their self-interests.
Second, after verifying the integrity of the deployed \textit{heavy/private} functions, we need to rebuild a connection between these functions and the already deployed \textit{light/public} functions (i.e., \textit{on-chain} contract). Since the two groups of functions have been separated from a single smart contract, such a connection is necessary to make them re-recognize each other, just as if they were deployed together in a single smart contract at the beginning.

\begin{figure}
\centering
{
    \includegraphics[width=8cm,height=5cm]{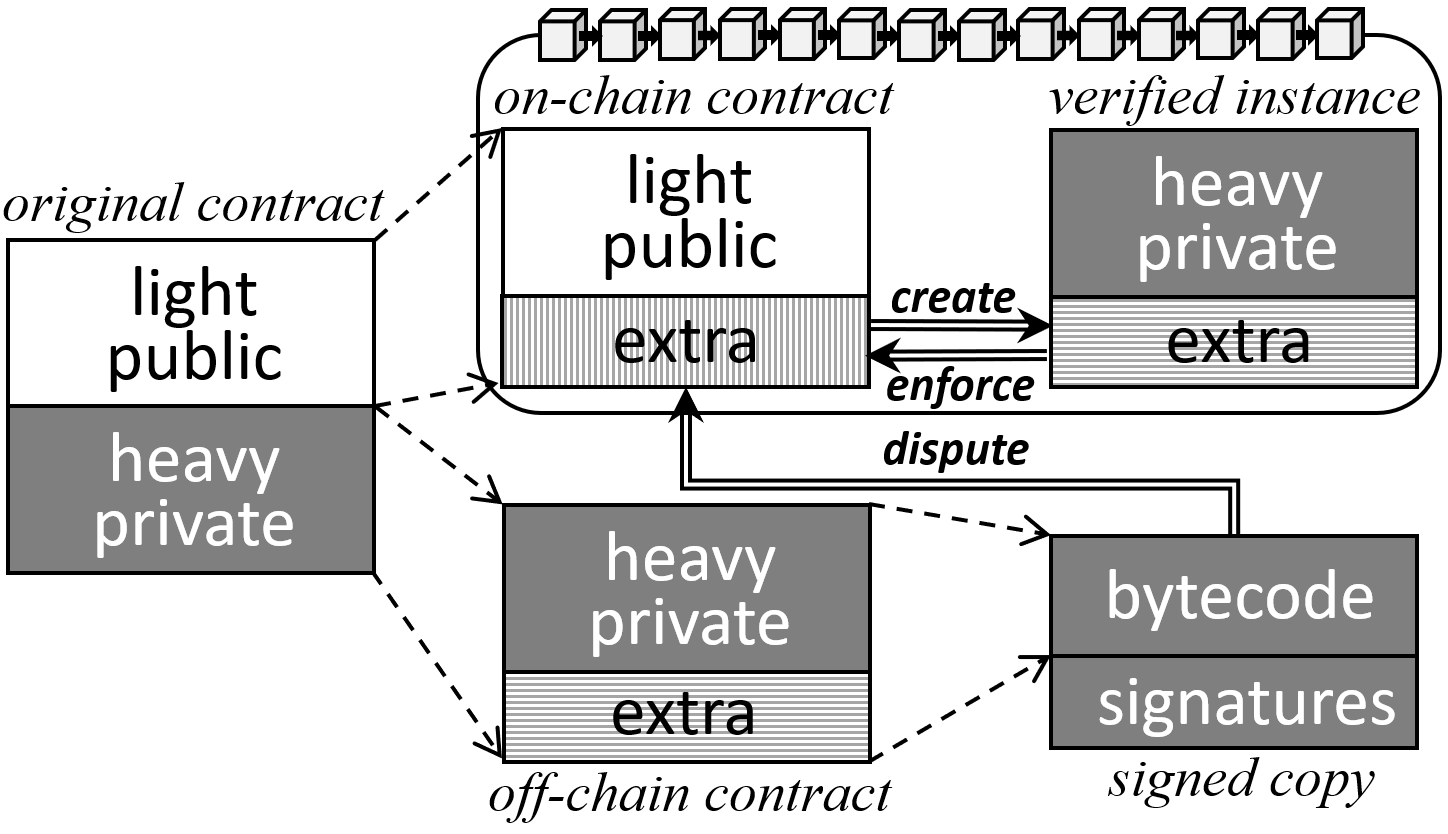}
}
\caption {\small Two categories of functions in the whole contract are padded with a few extra functions to form the \textit{on-chain} contract and the \textit{off-chain} contract. 
In the presence of dishonest participants, any honest participant can use the signed copy of the \textit{off-chain} contract to make the \textit{on-chain} contract create a verified instance of the \textit{off-chain} contract and leverage the unique link between the instance and its creator (i.e., \textit{on-chain} contract) to enforce state change of the \textit{on-chain} contract.
}
\label{f2} 
\vspace{-5mm}
\end{figure}

To overcome the aforementioned challenges altogether, in Fig.\ref{f2}, we propose a general mechanism of enforcing \textit{off-chain} contracts in the \textit{hybrid-on/off-chain} execution model of smart contracts, which consists of four stages:
\begin{itemize}[leftmargin=*]
\item \textit{Split/generate}:
Before any on-chain activity, the \textit{original} contract is first split to two parts that package \textit{light/public} functions and \textit{heavy/private} functions, respectively. 
Obviously, the two parts themselves usually do not contain any function for resolving a dispute, so we need to pad each group of functions with a few extra functions prepared for a dispute.
After padding, we have already generated an \textit{on-chain} contract as well as another \textit{off-chain} contract from the whole contract.
\item \textit{Deploy/sign}:
The \textit{on-chain} contract (i.e., \textit{light/public} functions including the padded extra functions) can then be deployed by any participant to the blockchain while the \textit{off-chain} contract (i.e., \textit{heavy/private} functions plus padded extra functions) needs to be converted into \textit{bytecode} to be signed by all the participants.
It is important to note that each participant must obtain a copy of the \textit{off-chain} contract with \textit{signatures} from all the participants before any interaction with the deployed \textit{on-chain} contract can take place. 
The procedure of generating \textit{signed} copies may easily be implemented through off-chain communication approaches, such as Whisper~\cite{Whisper2017} in Ethereum.
\item \textit{Submit/challenge}:
When all participants are honest, they can execute computation of the \textit{off-chain} contract by themselves and manually submit the results to the \textit{on-chain} contract to push the state change.
This can be implemented by leaving a challenge period after a representative of the participants have submitted the result, during which all other participants can challenge the result with the signed copy of the \textit{off-chain} contract.
Obviously, if the representative is honest, there will be no need to challenge the submitted result, so the state of the  \textit{on-chain} contract can be successfully changed as expected without revealing any information of the \textit{off-chain} contract and also without requiring miners to perform the computation of \textit{heavy/private} functions.
\item \textit{Dispute/resolve}:
In the presence of any dishonest participants, such as if the representative violates the agreement, a dispute occurs.
To resolve the dispute, during the challenge period, any honest participant can submit the \textit{signed} copy, namely the \textit{bytecode} and \textit{signatures} together, to the \textit{on-chain} contract, where an extra function padded to the \textit{on-chain} contract during the \textit{split/generate} stage will be called to verify the \textit{signed} copy through \textit{signatures} and create a verified \textit{on-chain} instance.
The deployed verified instance is created through the \textit{bytecode} of the \textit{off-chain} contract, so it is cloned from the \textit{off-chain} contract and consists of all the \textit{heavy/private} functions and padded extra functions in the \textit{off-chain} contract.
A participant (may or may not be the one submitting the \textit{signed} copy) can then invoke the \textit{heavy/private} functions within the verified instance to make them be executed by the miners. 
The execution result will then be the \textit{true} one, which is different from the \textit{false} result submitted by the dishonest participant during the \textit{dispute/resolve} stage.
Then, an extra function padded to the \textit{off-chain} contract will be invoked to send the \textit{true} result back to the \textit{on-chain} contract and another extra function padded to the \textit{on-chain} contract will receive the \textit{true} result, enforce the dispute resolution and penalize the dishonest participant. 
To sum up, in the fourth stage, three extra functions are designed to resolve a dispute:
\begin{itemize}[leftmargin=*]
\item \textit{deployVerifiedInstance()}: 
We design this extra function at the \textit{on-chain} contract to receive the \textit{signed} copy during the challenge period. Once called, this function first verifies all the received \textit{signatures} and then creates an \textit{on-chain} instance for the \textit{bytecode} within a \textit{signed} copy that has passed the verification.
\item \textit{returnDisputeResolution()}: 
We design this extra function at the \textit{off-chain} contract to return the \textit{true} result computed by miners to the  \textit{on-chain} contract.
\item \textit{enforceDisputeResolution()}:
We design this extra function at the \textit{on-chain} contract to enforce the dispute resolution.
\end{itemize}

\end{itemize}

\begin{table}
\begin{minipage}[h]{\linewidth}
\begin{small}
\begin{mdframed}[innerleftmargin=8pt]
\centerline{\textbf{Betting Rules}}
\textbf{Participants} Alice and Bob.

\noindent \textbf{Rules:} 
\begin{packed_enum}[leftmargin=*]
  \item Before time $T_0$, Alice should deploy \textit{on-chain} contract. Both the participants should keep a signed copy of the \textit{off-chain} contract.
  \item Before time $T_1$, both the participants can make a deposit (1 Ether) to \textit{on-chain} contract through \textit{deposit()} or request a refund through \textit{refundRoundOne()}.
  \item Between $T_1$ and $T_2$, if either Alice's balance or Bob's balance in \textit{on-chain} contract is not 1 Ether, the participants can request a refund through \textit{refundRoundTwo()}.
  \item The time point $T_2$ is the temporal threshold when the betting result becomes available. After that, between $T_2$ and $T_3$, the participants can compute the result \textit{off-chain}. The loser (say Alice) can then call \textit{reassign()} to make \textit{on-chain} contract transfer 2 Ether to winner's account.
  \item Finally, if the loser refused to implement step 4, the game will go to \textit{dispute/resolve} stage. After $T_3$, the winner (say Bob) can call the extra function \textit{deployVerifiedInstance()} with the signed copy at the \textit{on-chain} contract and then call the extra function \textit{returnDisputeResolution()} at the verified instance to enforce the dispute resolution.
\end{packed_enum}
\end{mdframed}
\end{small}
\vspace{-3mm}
\captionof{table}{Betting rules}
\vspace{-9mm}
\label{rules}
\end{minipage}
\end{table}

With this four-stage mechanism, honest participants always keep the ability to make dishonest participants be verified and monetary penalized, which in turn incentivizes all participants to honestly follow the results of off-chain smart contracts.
Next, we design and implement the proposed approaches using a detailed example.


\section{Implementation}
\label{s4}
We implement the proposed approaches using an example smart contract for betting between Alice and Bob. 
The betting rules in the example smart contract are listed in Table~\ref{rules}. Note that the presented betting rules do not involve penalty to dishonest participants, however it is straight-forward to revise step 2, 3 and 4 to monetarily penalize any dishonest participant in step 5.

The contracts are programmed with \textit{Solidity} and the \textit{off-chain} signature part is implemented using \textit{JavaScript} with ethereumjs-util~\cite{ethereumjs-util} package and web3-utils~\cite{web3-utils} package.
All the contracts have been tested over the Ethereum official test network \textit{Kovan}~\cite{Kovan2019}.

\vspace{2mm}

\begin{minipage}{\linewidth}
\begin{lstlisting}[
linewidth=8.3cm,
language=Solidity,
basicstyle=\scriptsize,
caption=\textit{on-chain} contract (interface),
label={a2},
frame=single,
numbers=none,
numbersep=5pt,
breaklines=true,
breakatwhitespace=true
]
pragma solidity ^0.4.24;

contract onChain {
  // light/public functions
  function deposit() payable public beforeT1 certifiedparticipantOnly;
  function refundRoundOne() public beforeT1 certifiedparticipantOnly;
  function refundRoundTwo() public T1toT2 certifiedparticipantOnly amountNotMet;
  function reassign() public T2toT3 certifiedparticipantOnly;
  // extra functions
  function deployVerifiedInstance(bytes memory _bytecode, uint8 _va, bytes32 _ra, bytes32 _sa, uint8 _vb, bytes32 _rb, bytes32 _sb) public afterT3 certifiedparticipantOnly amountMet;
  function enforceDisputeResolution(bool _winner) external deployedAddrOnly;
}
\end{lstlisting}
\end{minipage}

\begin{minipage}{\linewidth}
\begin{lstlisting}[
linewidth=8.3cm,
language=Solidity,
basicstyle=\scriptsize,
caption=\textit{off-chain} contract (simplified),
label={a3},
frame=single,
numbers=none,
numbersep=5pt,
breaklines=true,
breakatwhitespace=true,
morekeywords={chaoli}
]
pragma solidity ^0.4.24;

// interface
contract onChain {
  function enforceDisputeResolution(bool _winner) external;
}

contract offChain {
  modifier certifiedparticipantOnly {...}
  // heavy/private functions
  function reveal() private returns(bool) {...}
  // extra functions
  function returnDisputeResolution(address _addr) public certifiedparticipantOnly {
    onChain C_on = onChain(_addr);
    C_on.enforceDisputeResolution(reveal());
  }
}
\end{lstlisting}
\end{minipage}

In rule 1, before $T_0$, the \textit{on-chain} contract should be deployed and the participants should have obtained \textit{signed} copy of the \textit{off-chain} contract. We show the implementation of the two contracts with \textit{Solidity} in Algorithm~\ref{a2} and Algorithm~\ref{a3}, respectively.
As can be seen, in this example, the \textit{on-chain} contract consists of four light/public functions and two extra functions while the \textit{off-chain} contract consists of one heavy/private function and one extra function.
Here we note that the parameters, constructor and modifiers (e.g., beforeT1, certifiedparticipantOnly) are omitted in Algorithm~\ref{a2}.

\vspace{2mm}

\begin{minipage}{\linewidth}
\begin{lstlisting}[
linewidth=8.3cm,
language=python,
basicstyle=\scriptsize,
caption=Signature v-r-s generation,
label={a4},
frame=single,
numbers=none,
numbersep=5pt,
breaklines=true,
breakatwhitespace=true,
morekeywords={chaoli}
]
var web3Utils = require('web3-utils')
var ethUtils = require('ethereumjs-util')
var _bytecode = '0x608060405234801...bab40029'
var h_bytecode = web3Utils.soliditySha3(_code)
var h_bytecode_hex=new Buffer(h_bytecode.slice(2),'hex')

var privkey_a = new Buffer('499...38c', 'hex')
var vrs_a = ethUtils.ecsign(h_bytecode_hex, privkey_a)
var _va = vrs_a.v
var _ra = vrs_a.r.toString('hex')
var _sa = vrs_a.s.toString('hex')

var privkey_b = new Buffer('360...572', 'hex')
var vrs_b = ethUtils.ecsign(h_bytecode_hex, privkey_b)
var _vb = vrs_b.v
var _rb = vrs_b.r.toString('hex')
var _sb = vrs_b.s.toString('hex')
\end{lstlisting}
\end{minipage}

\begin{minipage}{\linewidth}
\begin{lstlisting}[
linewidth=8.3cm,
language=Solidity,
basicstyle=\scriptsize,
caption=\textit{deployVerifiedInstance()},
label={a5},
frame=single,
numbers=none,
numbersep=5pt,
breaklines=true,
breakatwhitespace=true
]
contract onChain {
  ...
  address public deployedAddr;
  
  function deployVerifiedInstance(bytes memory _bytecode, uint8 _va, bytes32 _ra, bytes32 _sa, uint8 _vb, bytes32 _rb, bytes32 _sb) public afterT3 certifiedparticipantOnly amountMet {
    // verify signatures
    bytes32 h_bytecode = keccak256(_bytecode);
    address a = ecrecover(h_bytecode, _va, _ra, _sa);
    address b = ecrecover(h_bytecode, _vb, _rb, _sb);
    require(a == certifiedparticipant[0] && b == certifiedparticipant[1]);
    // create verified instance
    address _addr;
    assembly {
      _addr := create(0, add(_code, 0x20), mload(_code))
    }
    deployedAddr = _addr;
  }
  ...
}
\end{lstlisting}
\end{minipage}

\begin{minipage}{\linewidth}
\begin{lstlisting}[
linewidth=8.3cm,
language=Solidity,
basicstyle=\scriptsize,
caption=\textit{enforceDisputeResolution()},
label={a6},
frame=single,
numbers=none,
numbersep=5pt,
breaklines=true,
breakatwhitespace=true
]
contract onChain {
  ...
  function enforceDisputeResolution(bool _winner) external deployedAddrOnly {
    accountBalance[participant[0]] = 0;
    accountBalance[participant[1]] = 0;
    if(_winner == true) {
      participant[1].transfer( accountBalance[participant[0]] + accountBalance[participant[1]]);
    } else {
      participant[0].transfer( accountBalance[participant[0]] + accountBalance[participant[1]]);
    } 
  }
  ...
}
\end{lstlisting}
\end{minipage}

After generating the two contracts, Alice and Bob should transform the \textit{off-chain} contract shown in Algorithm~\ref{a2} into \textit{bytecode} through tools such as \textit{Remix} or \textit{Truffle} and then use the \textit{JavaScript} program in Algorithm~\ref{a4} to generate signatures (tuples of \textit{(v,r,s)}) with their account private keys and the \textit{bytecode}.
Please note that all the participants should use the same version of compiler for the purpose of getting same \textit{bytecode}.

In case if Alice or Bob violates the rules before $T_2$, the game can always be terminated through \textit{refundRoundOne()} or \textit{refundRoundTwo()}.
After $T_2$, suppose Alice loses the game, she should call \textit{reassign()} to admit this failure before $T_3$ to let the function transfer both winner's deposit and loser's deposit to winner.
If \textit{reassign()} is not called before $T_3$, the winner will find the fact that the loser has violated the rules and a dispute has occurred.
Once the dispute happens, after $T_3$, the winner (say, Bob in this example) should call the \textit{deployVerifiedInstance()} extra function (Algorithm~\ref{a5}) with the \textit{signed} copy. 
The extra function will first verify the signatures by outputting an address from \textit{bypecode} and \textit{(v,r,s)} and checking whether it is the signer's address.
After that, the extra function will create a verified instance from the \textit{bytecode} and also record the address of the created instance in the parameter \textit{deployedAddr}.
Thus, by generating the verified instance from the \textit{on-chain} contract, the verified instance can be authorized through the parameter \textit{deployedAddr} because no other contract can have the same address and it is also guaranteed that the verified instance has been agreed by all the participants.
Please note that to create a contract from an existing contract with only \textit{bytecode}, we must use the assembly language in \textit{Solidty}.

Finally, the winner (i.e., Bob) can call \textit{returnDisputeResolution()} in the verified instance (see Algorithm~\ref{a3}) with the address of the \textit{on-chain} contract (see Algorithm~\ref{a2}).
The input address of the \textit{on-chain} contract will then allow the verified instance to return the dispute resolution to the \textit{on-chain} contract by invoking the extra function \textit{enforceDisputeResolution()} (Algorithm~\ref{a6}), which then enforces both winner's deposit and loser's deposit to be transferred to winner and may also monetarily penalize the dishonest loser if there is a penalty rule.
Here, the \textit{deployedAddrOnly} modifier will check \textit{msg.sender} so that only the verified instance with the address same as the one recorded in parameter \textit{deployedAddr} can leverage \textit{enforceDisputeResolution()} to change the state of \textit{on-chain} contract.

\begin{table}[h]
\centering
\begin{tabular}{|c |c |} \toprule 
{\textbf{Extra function}} & {\textbf{Gas cost}} \\ \midrule
    \textit{deployVerifiedInstance()} & 225082 + \textit{reveal()} \\
    \midrule
    \textit{returnDisputeResolution()} & 37745 \\
    \bottomrule
\end{tabular}
\vspace{-1mm}
\caption{Gas cost}
\label{gas}
\vspace{-3mm}
\end{table}

We now present the gas cost for dispute resolution. 
As shown in Table~\ref{gas}, in case of a dispute, an honest participant needs to first spend (225082 + cost of \textit{reveal()}) gas to deploy the verified instance through \textit{deployVerifiedInstance()} and then spend 37745 gas to enforce the resolution through \textit{returnDisputeResolution()}. 
The overall cost is not high when cost of \textit{reveal()} is low.
However, if \textit{reveal()} is a heavy function, it should be mandatory for each participant to pay security deposit so that the honest participant paying for dispute resolution can receive compensation from dishonest participants.

\section{Related work}
\label{s5}

Recently, \textit{off-chain} resources of blockchain have been widely studied by researchers to improve the performance of blockchains~\cite{eberhardt2017or,kalodner2018arbitrum,li2018decentralized2,li2018decentralized,molina2018implementation,poon2016bitcoin}. 
Among them, \cite{kalodner2018arbitrum} and~\cite{molina2018implementation} are most relevant to our work. 
In~\cite{molina2018implementation}, smart contracts are implemented using hybrid architectures similar to the \textit{hybrid-on/off-chain} execution model proposed in our work. However, the proposed hybrid architectures in ~\cite{molina2018implementation} rely on a Trusted Third Party (TTP) as an oracle and thus the architectures are not completely decentralized.
In~\cite{kalodner2018arbitrum}, a smart contract system named Arbitrum is developed, where smart contracts are designed to be executed \textit{off-chain}. As the system is specially designed for this purpose, it is hard to generalize the system-level design to existing systems such as Ethereum. 
In this paper, instead of treating the use of \textit{off-chain} contracts as a system-level design goal, we consider the \textit{hybrid-on/off-chain} computation model as an application-level smart contract design pattern and also as a building block for enhancing blockchain scalability and privacy. Thus, the proposed approach is a plug-and-play solution that is compatible with existing smart contract systems and their time-tested infrastructure and community.
In addition, the combination of the proposed approach and other system-level or application-level solutions, such as sharding~\cite{al2017chainspace} and zero knowledge proof~\cite{kosba2016hawk}, can further enhance the scalability and privacy of the smart contract systems.

\section{Conclusion}
\label{s6}
In this paper, we propose a general \textit{hybrid-on/off-chain} execution model of smart contracts, which separates the heavy/private functions of a smart contract from the light/public ones to form an \textit{off-chain} contract and enables the \textit{off-chain} contract to be executed only by the interested participants. The proposed approach leads to increased saving of computing resources of the miners and protects sensitive information of a smart contract in the blockchain.
To handle disputes caused by any dishonest participants, we propose a mechanism that allows any honest participant to reveal a signed copy of the \textit{off-chain} contract so that a verified instance can be created to make miners enforce the true execution result.
Therefore, with the proposed approach, honest participants always have the ability
to redress and penalize any fraudulent or dishonest behavior, which incentivizes all participants to honestly follow the agreed \textit{off-chain} contract.
Our implementation and evaluation of the proposed approach using
an example smart contract demonstrate the effectiveness of the
proposed approach of using \textit{on/off-chain} contracts in Ethereum.

\renewcommand\refname{Reference}
\bibliographystyle{plain}
\urlstyle{same}
\bibliography{main.bib}

\end{document}